\documentclass[preprint,superscriptaddress,showpacs,amsmath,amssymb,aps,pre,floatfix]{revtex4-1}
\usepackage[dvipsnames,usenames]{color}
\usepackage{graphicx, subfigure}
\usepackage{rotating}
\usepackage{dcolumn}
\usepackage[normalem]{ulem} 
\usepackage{hyperref}
\usepackage{nameref}
\usepackage{morefloats}

\graphicspath{{figs_v4/}}

\begin{document}

\title{Uncovering the role of spatial constraints in the differences and similarities between physical and virtual mobility}

\newcommand{\RochesterP}{Department of Physics \& Astronomy, University of Rochester, Rochester, NY, USA}
\newcommand{\Exeter}{Department of Computer Science, University of Exeter, Exeter, UK}

\author{Surendra Hazarie} \affiliation{\RochesterP}
\author{Hugo Barbosa} \affiliation{\RochesterP} 
\affiliation{\Exeter}
\author{Adam Frank} \affiliation{\RochesterP}
\author{Ronaldo Menezes} \affiliation{\Exeter}
\author{Gourab Ghoshal} \affiliation{\RochesterP}

\begin{abstract} 

The recent availability of digital traces from Information and Communications Technologies (ICT) has facilitated the study of both individual- and population-level movement with unprecedented spatiotemporal resolution, enabling us to better understand a plethora of socioeconomic processes such as urbanization, transportation, impact on the environment and epidemic spreading to name a few. Using empirical spatiotemporal trends, several mobility models have been proposed to explain the observed regularities in human movement. With the advent of the World Wide Web, a new type of virtual mobility has emerged that has begun to supplant many traditional facets of human activity. Here we conduct a systematic analysis of physical and virtual movement, uncovering both similarities and differences in their statistical patterns. The differences manifest themselves primarily in the temporal regime, as a signature of the spatial and economic constraints inherent in physical movement, features that are predominantly absent in the virtual space. We demonstrate that once one moves to the time-independent space of events, i.e the sequences of visited locations, these differences vanish, and the statistical patterns of physical and virtual mobility are identical. The observed similarity in navigating these markedly different domains point towards a common mechanism governing the movement patterns, a feature we describe through a Metropolis-Hastings type optimization model, where individuals navigate locations through decision-making processes resembling a cost-benefit analysis of the utility of locations. In contrast to existing phenomenological models of mobility, we show that our model can reproduce the commonalities in the empirically observed statistics with minimal input.

\end{abstract}

\maketitle

\date{}
%
%


\section{Introduction}

Recent years have witnessed an explosion of extensive geolocated datasets related to human movement, enabling scientists to quantitatively study individual and collective mobility patterns, and to generate models that can capture and reproduce the spatiotemporal structures and regularities in human trajectories~\cite{Barbosa-Filho2017}. The study of human mobility is especially important for applications such as estimating migratory flows, traffic forecasting, urban planning, mitigating pollution and epidemic modeling~\cite{Batty2013,simini_2012_universal,Uherek2010,Lee_2017,Kirkley_2018,Pan_2013,tizzoni_2012_real,toole_2015_path}. A particularly rich source of data has been from geotagged traces, including Call Data Records (CDR's)\cite{Blondel_2015_CDRreview} and location based social networking services (LSBN's)~\cite{Barbosa2015}. Several common regularities have been observed across these studies, including bursty activity rates~\cite{Vazquez2006}, tendencies to visit a select few locations disproportionately more than others, as well as decreasing likelihood to explore as time goes on~\cite{Gonzalez2008a}. Based on these findings, a series of phenomenological models have been proposed to explain the observed regularities~\cite{Brockmann2006, Song2010,Rhee2011a, Boyer2012, Hasan2013a}.

 
However, with the advent of the World Wide Web, it is possible to define and study an entirely new dimension of human mobility, that is, in the virtual space, a feature that has been rapidly gaining attention~\cite{Radicchi_2009, Zhao_2016,Hu_2018, Hu_2019}. This is of course natural, given that an increasing fraction of human activities such as  shopping, information consumption, education, and social interaction are being replaced by their online counterparts. This phenomenon is relatively recent (on a biological time-scale) and much remains to be uncovered; therefore, a better understanding of the long-term impacts of such changes in behavior and the corresponding challenges is crucial.

Increasing evidence suggests that online activity, including virtual mobility, is governed by similar mechanisms influencing offline activity. For instance, foraging behaviors have been  observed in contexts as diverse as online shopping \cite{Hantula2008} and Web searching \cite{Wang2017}, with striking similarities to primitive resource acquisition behavior~\cite{stephens_2007_foraging}. Furthermore, many statistical regularities observed in physical movement have also been observed in virtual movement, including the distribution of visitation frequencies to locations~\cite{Zhao2014}, power-law distributions of activity rates in on-line bookmarking~\cite{Wang2011}, and in the special case of virtual worlds, the heavy-tailed distribution of displacements~\cite{Szell2012} (a feature that is unreproducible in most online movement, due to the lack of a metric space). Even more strikingly, a method of characterization of individuals according to whether they have exploratory or saturation behavior in terms of location discovery, has been successfully adapted to Web browsing~\cite{Cherifi2016}, leading to similar findings as those obtained from the analyses of physical trajectories \cite{Pappalardo2015}.

At first blush, the observed similarity in mobility trends in physical and virtual domains is rather surprising. In particular, physical movement is necessarily constrained by temporal and economic costs in terms of moving from one location to the other. For instance, it is well known that resident movement in cities is shaped by elements such as wealth distribution~\cite{Lotero2014} as well as transportation hierarchy~\cite{Gallotti2015}. Yet, no such limitations exist while navigating the Web, which is neither economically (for the most part) nor spatially bounded. Furthermore, despite the recent spate of research on online activities, there are relatively fewer studies conducting a direct comparison between physical and virtual mobility. While much has been said about their commonalities, less attention has been devoted to their differences, specifically, the role played by the inherent spatial costs in physical movement.

To fill this gap, in this paper, we conduct a systematic analysis of the similarities and differences between online and offline movement. We study one virtual and two physical datasets, finding that the differences arising from the cost associated with spatial movement, manifest as differences in temporal mobility statistics primarily at shorter time-scales corresponding to the intra-day regime. Once we move to the time-independent space of events, that is sequences of location visits, these differences dissipate, and the statistical patterns are essentially indistinguishable, pointing to a common mechanism underlying both behaviors. 

Inspired by research on human decision-making and optimization strategies for foraging, we propose a generative model based on the Metropolis-Hastings algorithm. Our model describes a population that traverses locations (both physical and virtual) as a function of maximizing the utility or benefit of moving to that location, balanced against the associated cost to do so. Using three simple features, a utility distribution corresponding to how the population perceives different locations, an exploration parameter that measures sensitivity to cost, and a cognitive-limit/incomplete-information parameter, that measures the number of assessed locations at any point in the decision-making process, we reproduce all of the empirically observed mobility trends. Unlike existing phenomenological models, our formulation requires minimal empirical input and is flexible enough to admit a wide spectrum of mobility patterns. We end with a discussion of the implication of our results.
\section{Results}

\subsection{Data}

For the purposes of our analysis, we consider user activities in one virtual dataset corresponding to Web browsing, and two physical datasets related to location-based social networking services. The browsing data consists of of $5.13 \times 10^6$ URL visitation events recorded by 521 anonymized users between September 2010 and May 2014~\footnote{\texttt{http://Webhistoryproject.blogspot.com}}. To more accurately reflect the correspondence between Web browsing and physical movement, the URLs are aggregated at the resolution of website hosts. This is to ensure that events corresponding to browsing a particularly complex website domain consisting of many individual URLs  do not introduce spurious attributes to the statistical patterns (akin to departments of a physical store being considered separately). Furthermore, we exclude redirections and embedded content, and only include events corresponding to explicit user requests, including link-clicks, bookmark-clicks, and explicit typing of URLs. 
 
 \begin{table*}[h!]
      \begin{minipage}{.5\linewidth}
           \centering
      \begin{tabular}{l|l|l}
User ID & Location & Time Elapsed (seconds) \\ \hline
1       & 1           & 1                      \\ \hline
1       & 14          & 20                     \\ \hline
1       & 20          & 40                     \\ \hline
1       & 7           & 45                     \\ \hline
\end{tabular}
    \end{minipage}%
    \begin{minipage}{.5\linewidth}
      \centering
   
 \begin{tabular}{l|l|l}
User ID & Host ID & Time Elapsed (seconds) \\ \hline
3       & 1           & 1                      \\ \hline
3       & 25          & 14                     \\ \hline
3       & 12          & 60                    \\ \hline
3       & 7           & 65                     \\ \hline
\end{tabular}

    \end{minipage} 
      \caption{{\bf Snapshots of user activity}  Users and locations/websites are assigned unique identifiers, and each event (corresponding to a location visit) is  timestamped corresponding to when that location was visited in the user's temporal history.} 
\label{tab:data}
\end{table*}

One of the physical mobility datasets was sourced from BrightKite, a location-based social networking service~\cite{Cho2011, Grabowicz2014}, containing $4.7 \times 10^6$ geotagged check-ins produced by approximately 51,400 users across fifty four months~\footnote{\texttt{https://snap.stanford.edu/data/loc-brightkite.html}}. The second dataset is from Weeplaces, a website that  generated visualizations and reports from location-based check-ins in platforms such as Facebook, Foursquare and Gowalla. The considered data contains only Foursquare check-ins, that include $7.7 \times 10^6$ geotagged check-ins produced by approximately 15,800 users across ninety two months~\footnote{\texttt{http://www.yongliu.org/datasets/.}}. 

Each \emph{event}, i.e an instance of a location visit, is timestamped and tagged with a unique location ID (physical) or website host ID (online). In order to filter out spurious activities such as automatic page refreshes or repeated app-specific check-ins, we exclude events that occur less than one second apart, as well as consecutive events at the same location. After the filtering procedure, we are left with $1.2 \times 10^6$ events across 519 users for Web-browsing, $2 \times 10^6$ events across 43,240 users in BrightKite, and finally $7\times 10^6$ events across 15,777 users for Weeplaces.  In all datasets a location visit $v$ is represented by a tuple $v = (u,\ell,t)$, meaning a user $u$ visited a location $\ell$ at time $t$. At a user level, a trajectory composed of $n_u$ discrete observations is characterized by a sequence of $n_u$ location-time pairs $(\ell_{i},t_{i})_{i \in 1\ldots n_u}$ where $\ell_{i}$ stands for the location visited at step $i$ and $t_{i}$, the timestamp in which such a visit has occurred. A typical snapshot of the dataset for both the virtual and physical mobility is shown in Tab.~\ref{tab:data}.

\subsection{Temporal statistics}

Ostensibly, one would expect physical and virtual movement to be different due to the inherent physical cost of traversing between locations in the former, a feature that is absent in the latter. Apart from the economic cost in terms of expending fuel or energy to traverse physical space, there is an associated temporal cost that far exceeds the time taken to traverse between Webpages. 

\begin{figure}[t!]
\includegraphics[width=\columnwidth]{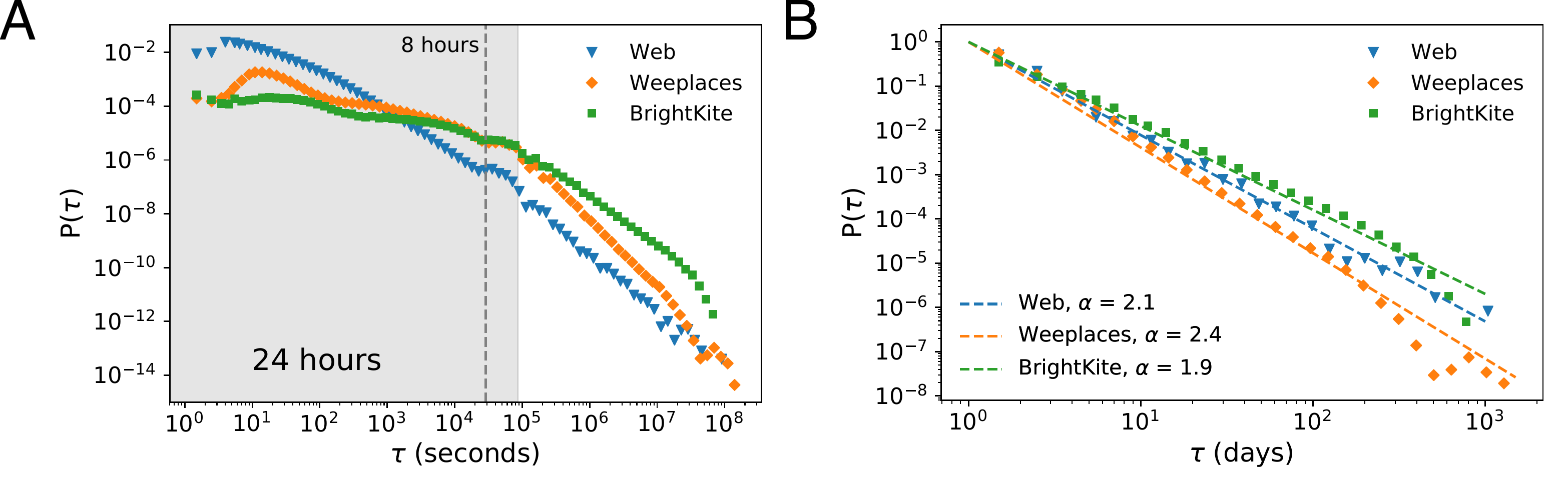}
  \caption{\textbf{Distribution of inter-event times at multiple temporal resolutions} \textbf{(A)} Eq.~\eqref{eq:iet_pop} plotted at the resolution of seconds, indicating noticeably more rapid activity in Web browsing as compared to moving between physical locations. The latter exhibits distinct scaling regimes at the intra- and inter-day levels, with a flatter distribution in the intra-day regime. A distinct bump is seen at $\tau \approx 8$ hours in all three datasets, corresponding to scheduled activities, circadian patterns or behavioral traits. These features disappear at the resolution of days \textbf{(B)} where all three datasets exhibit a scaling of the form $P(\tau) \sim \tau^{-\alpha}$ (fits to the data using a maximum likelihood estimate~\cite{Clauset2009}, shown as dashed lines).}
     \label{fig:int_event}
\end{figure}

\subsubsection{Inter-event times}
We begin our analysis of this feature, by calculating the distribution of inter-event times in each of the datasets~\cite{Radicchi_2009}.  For a given user $u$, initiating $n_u$ events in their temporal history $t_1 \ldots t_n$, the inter-event time is defined as $\tau^{u}_i = \left(t_{i+1} - t_i\right)$. From this, one can calculate the probability that two events initiated by user $u$ are separated by time $\tau$ thus,
\begin{equation}
P_u(\tau) = \frac{1}{n_u -1} \sum_{i = 1}^{n_u-1} \delta_{\tau,\tau^{u}_i},
\label{eq:iet_u}
\end{equation}
where $\delta_{i,j}$ is the Kronecker delta. Finally, the population distribution of the inter-event time is given by,
\begin{equation}
P(\tau) = \frac{\sum_{u = 1}^{N_u} (n_u -1) P_i(\tau)}{\sum_{u =1}^{N_u} (n_u -1)},
\label{eq:iet_pop}
\end{equation}
where $N_u$ represents the total number of users. 

We plot $P(\tau)$ for the three datasets in Fig.~\ref{fig:int_event}, at the resolution of seconds and days, finding a broad distribution in all cases; short-intervals between activities are far more common than longer ones, although virtual mobility exhibits significantly more rapid events then physical mobility at the resolution of seconds (Fig.~\ref{fig:int_event}A), with the difference diminishing at the resolution of days (Fig.~\ref{fig:int_event}B). This is indicative of the greater cost in physical movement, whereby rapid transitions between locations are necessarily constrained by the limitations of transportation which is temporally more costly than clicking on links. Yet, in all three cases, we see a relative increase in probability around eight hours, indicative of events as parts of daily routines. For instance, the time elapsed between the final event in the day (arriving home or visiting social media before bed) and the first event of the next day (leaving for work or school or checking emails), or from scheduled activities (work routines, checking email). We note that these effects are present regardless of whether the activities are virtual or physical, and correspond to circadian rhythms and consistent behavioral patterns.

However, there is a marked difference in the temporal patterns between physical and virtual mobility suggesting characteristic time-scales associated with each type of activity. The location-based social networks exhibit two distinct regimes of behavior corresponding to the intra-day range ($\tau \leq 24$ hours) where the distribution is relatively flat as compared to the inter-day range ($\tau > 24$ hours), a feature that is absent in virtual browsing. This difference likely arises from the fact that daily physical schedules contain extended periods of time at single locations such as work or school, a behavior that is not typical of online browsing. Small alterations to schedules (such as offsite lunches) turn these previously uninterrupted events into trajectories of smaller events---a work day split into before-during-and after lunch events---that flatten the intra-day regime of the physical mobility data. In the inter-day regime, all three datasets follow a similar power-law scaling of the form $P(\tau) \sim \tau ^ {-\alpha}$ where $\alpha_{\textrm{Web}} \approx 2.1$, $\alpha_{\textrm{Weeplaces}} \approx 2.5$ and $\alpha_{\textrm{BrightKite}} \approx 1.9$.   

The results indicate that the primary difference in activity rates between virtual and physical mobility is at shorter time-scales, where the cost associated with physical space plays a primary role. Once one examines longer time-scales, the differences vanish, suggesting that cognitive processes, human behavioral traits, and daily schedules play a larger role in shaping the observed patterns.

\begin{figure*}[t!]
\includegraphics[width=\columnwidth]{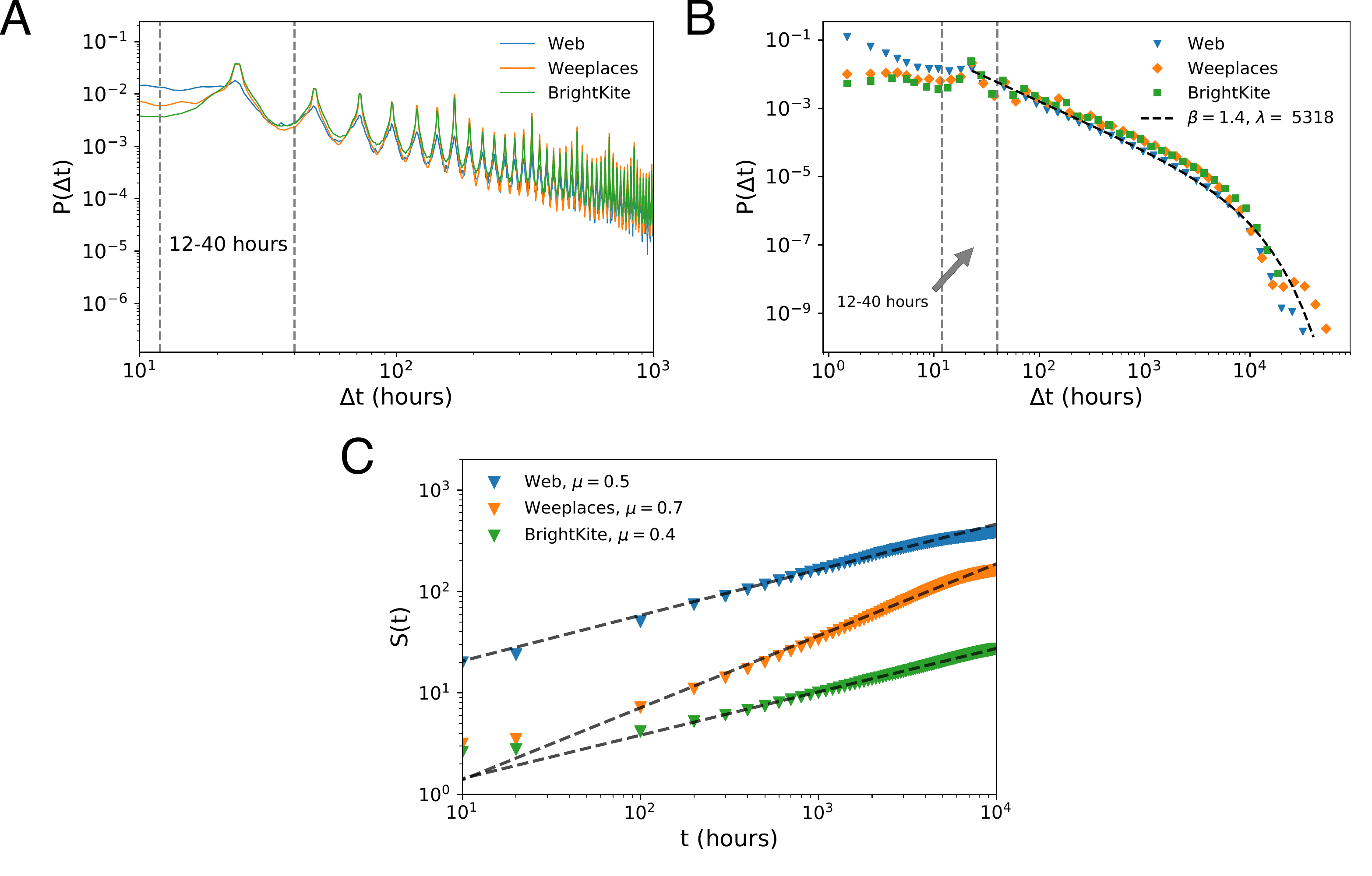}
  \caption{\textbf{Temporal Statistics} \textbf{(A)} The distribution of inter-return times (Eq.~\eqref{eq:irt_pop}) plotted with linear bins. Each peak corresponds to integer-days, and all three datasets indicate a clear circadian pattern. \textbf{(B)} Inter-return time distributions plotted with logarithmic bins. All three datasets follow a truncated power-law form $p(\Delta t) \sim \Delta t^{-\beta} \exp(-\Delta t /\lambda)$ shown as dashed curve. \textbf{(C)} The number of discovered unique locations as a function of time $S(t)$(Eq.~\eqref{eq:st}). While, all datasets scale sub-linearly $S(t) \sim t^{\mu}$ (fits shown as dashed curves), an order of magnitude separates the number of discovered locations in the Web as compared to physical movement. The rate of discovery (indicated by the slope) is roughly the same for the Web and BrightKite, while it is relatively more rapid in Weeplaces.}
\label{fig:return_sn}
\end{figure*}

\subsubsection{Inter-return times}
Indeed, one of the observed signatures of behavioral traits and realization of daily schedules is the tendency to return to the same location. This regularity has been observed both in physical and virtual movement, and has been used to partition the population based on returning and exploratory tendencies~\cite{Tauscher1997, Szell2012, Gonzalez2008a, Cherifi2016,Pappalardo2015}. One way to capture this regularity, is to measure the inter-return time, the time elapsed between visits of an individual to the same physical or virtual location. More precisely, for a given user $u$ visiting location $\ell$, we calculate the inter-return-times between visits $i$ and $i+1$ to $\ell$ as $\Delta t^u_{\ell,i} = t^u _{\ell,i+1} - t^u _{\ell,i}$. Then, the probability that consecutive visits to the same location are separated by a time $\Delta t$, for total visits $k_{\ell}$ to location $\ell$, summed over all locations is,
\begin{equation}
    P_u(\Delta t) = \frac{1}{\sum_{\ell} k_{\ell} - 1} \sum_{\ell} \sum_i^{k_{\ell}-1} \delta_{\Delta t, \Delta t^u_{\ell,i}},
\end{equation}
with the distribution over the population given by 

\begin{equation}
P(\Delta t) = \frac{\sum_{u = 1}^{N_u} (n_u -1) P_i(\Delta t)}{\sum_{u =1}^{N_u} (n_u -1)}.
\label{eq:irt_pop}
\end{equation}


We plot the inter-return time distributions in Fig.~\ref{fig:return_sn} A at the resolution of hours, finding remarkably similar periodic behavior across three datasets. The observed peaks are indicative of circadian rhythms, corresponding to integer-day returns such as visits to home, work or weekly trips to stores. For physical movement, these trends are expected features being natural consequences of the physical constraints of daily life. However, these constraints are completely absent in virtual mobility, and yet remarkably, these trends persist.  Plotting the distributions in coarser logarithmic bins in panel B, we see that all three datasets follow a truncated power-law form $P(\Delta t) = \Delta t ^ {-\beta} \exp\left(-\Delta t/\lambda\right)$ with the same scaling exponent $\alpha \approx 1.4$ and cut-offs $\lambda_{\textrm{Web}} = 4047$ hours, $\lambda_{\textrm{Weeplaces}} = 7092$ hours, and $\lambda_{\textrm{BrightKite}} = 4813$ hours. Similar to what is seen for the inter-event times, we do note a larger amount of intra-day returns for Web browsing (Fig.~\ref{fig:return_sn}B), but going beyond 24 hours, the scaling and cutoffs are similar across physical and virtual activities, indicating a shared upper limit to the time that humans are willing to revisit locations. 

\subsubsection{Temporal discovery rate}

Having characterized the returning tendencies, we next investigate the differences in exploratory patterns between the two types of movement. In the physical context, exploration can be cast as the characteristic distance one travels from one's base location (home or office for instance). Such an analysis has been previously conducted on mobile phone data using the so-called radius of gyration---a measure of the aforementioned characteristic distance---which was found to increase logarithmically in time, indicating strong saturation limits to the distance people travel from a source location~\cite{Gonzalez2008a, Song2010}. The observed saturation is a natural consequence of the temporal, physical and economic costs associated with movement. It is difficult to define an analog for the radius of gyration in the context of virtual browsing, given that there is no associated metric space. Instead, to examine exploratory tendencies in a way that applies to both physical and virtual space, we calculate the number of unique locations $S_u$ visited by user $u$ throughout the extent of the observation period~\cite{Tauscher1997}.  We define the number of unique elements in a user $u$'s trajectory at time $t$ as $S_{u,t}$, and calculate the average across all $N$ users thus,
\begin{equation}
 S(t) = \frac{\sum_u S_{u,t}}{N}.
 \label{eq:st}
\end{equation} 
In Figure \ref{fig:return_sn}C we show $S(t)$ for all three datasets finding in each case a sub-linear scaling of the form~$S(t) \sim t^{\mu}$ where $\mu_{\textrm{Web}} \approx 0.5$, $\mu_{\textrm{Weeplaces}} \approx 0.7$, and $\mu_{\textrm{BrightKite}} \approx 0.4$. The results indicate saturation effects in Web browsing as well, however, unlike returning tendencies, where apart from intra-day activities, physical and virtual movement are essentially indistinguishable, the number of unique locations visited in the Web is an order of magnitude higher than in the physical datasets. This is likely due to the combination of two factors; the fact that there are simply many more locations in the virtual realm than in the physical one, and second, the ability to sample such locations is not constrained by physical limitations. Instantaneous browsing of websites implies that individuals will naturally discover new locations faster than they could physically.  It is instructive to note, however, that while there is a marked difference in the number of locations visited, the rate of discovery, reflected in the slope of $S(t)$, is virtually identical for both the Web and BrightKite. We note the relatively more rapid discovery rate found in Weeplaces; this is likely due to the fact that in Foursquare checkins, users are actively incentivized to discover new locations. 

\subsection{Time-independent statistics}

The temporal statistics examined thus far, indicate that the putative restrictions on physical mobility as compared to virtual mobility, manifest themselves as relatively more rapid activity at shorter time-scales and the tendency to visit significantly more locations in a comparable time-window for the latter. Yet the disappearance of these differences at longer time-scales and the similarities in the discovery rates hint at a common mechanism driving the mobility behavior in both domains. To investigate this, we next study those aspects of mobility that are time-independent. 

\subsubsection{Event discovery rate}

\begin{figure}[t!]
\includegraphics[width=\columnwidth]{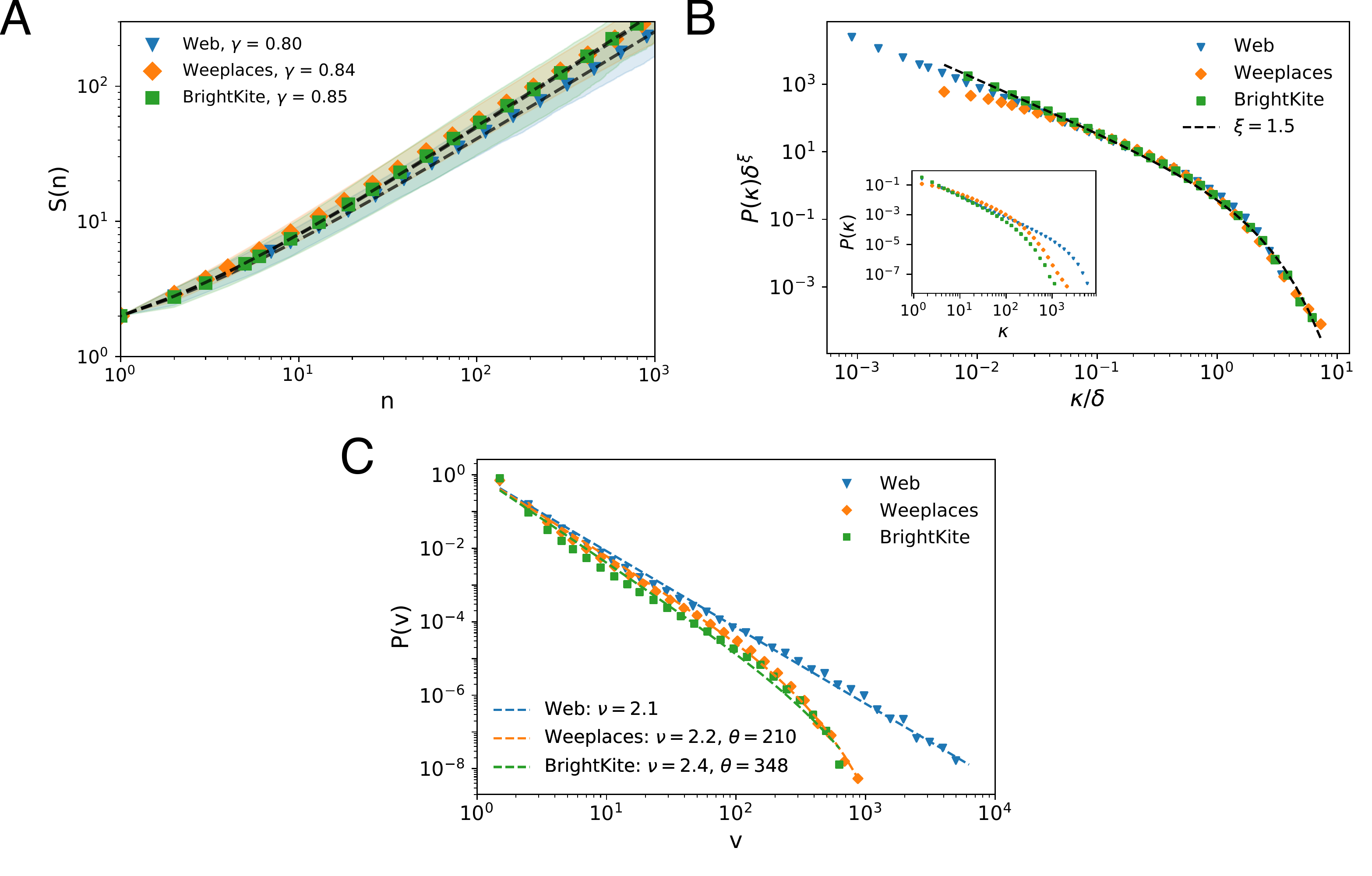}
  \caption{\textbf{Time-independent Metrics} \textbf{(A)} The number of unique locations visited as a function of event count $S(n)$ (Eq.~\eqref{eq:sn}). Unlike for $S(t)$, the trends in all three datasets are similar, whereby $S(n) \sim n^{\gamma}$ with $\gamma \approx 0.8$. Fluctuations over the population shown as shaded region. \textbf{(B)} The recency effect is present in both the virtual and physical domains, with the distribution of unique intermediate locations visited, before returning to a given location $P(\kappa)$ (Eq.~\eqref{eq:pkappa}) following a truncated power-law distribution, $P(\kappa) \sim \kappa^{-\xi} \exp\left(-\kappa/\delta \right)$, with a common scaling exponent  $\xi \approx 1.5$ (inset). After rescaling, with respect to the cut-offs, all three distributions collapse on to the same curve (shown as dashed line).\textbf{(C)} The frequency distribution of location visits for all three datasets follow the form $P(v) \sim v^{-\nu}$ with roughly similar exponents, however we note the presence of exponential cut-offs $\exp \left(-v/\theta \right)$ in the physical mobility distribution.}
     \label{fig:serial_metrics}
\end{figure}

To examine exploration in virtual and physical space without the bias of temporal cost, we define the number of unique locations visited in a user $u$'s trajectory at event $n$ as $S_{u,n}$ and calculate the population average thus,
\begin{equation}
    S(n) = \frac{\sum_u S_{u,n}}{N}.
    \label{eq:sn}
\end{equation}
The quantity $S(n)$ measures discovery as a function of the number of initiated events, and is the time-independent analog of $S(t)$. We plot $S(n)$ in Fig.~\ref{fig:serial_metrics} A, finding that the differences between physical and virtual exploration are negligible. While the exploration rate continues to be sub-linear $S(n) \sim n^{\gamma}$, ($\gamma < 1)$, unlike the temporal discovery rate, the scaling exponent is roughly comparable across the three datasets: $\gamma_{\textrm{Web}} \approx 0.8 $, $\gamma_{\textrm{Weeplaces}} \approx 0.85$, $\gamma_{\textrm{BrightKite}} \approx 0.85$. Furthermore, the observed stark differences in the number of unique locations visited between the Web and physical datasets also dissipate, suggesting that exploratory tendencies as measured by initiated events are the same in both domains. Alternatively, each transition in physical mobility is just as likely to discover a new location as a corresponding transition in virtual mobility is to discover a new Web host. The continued presence of sub-linear scaling in the event space, indicates that bounds on exploration are perhaps more reflective of cognitive limits, as opposed to physical constraints. 

\subsubsection{Recency effect}

Next, we seek the time-independent analog of the returning tendencies. A natural way to quantify this is the so-called recency effect, the observed tendency of individuals to better recall immediate items in a sequential list~\cite{ebbinghaus1913memory}.  The recency effect has been shown not only to influence memory, but also decision making processes. For instance, while evaluating figure skaters, those who perform later (and therefore more recently in the judge's mind) are regarded more favorably~\cite{BruinedeBruin2005}. Similar patterns were reported in singing competitions~\cite{Li2009} and in humanitarian aid allocation \cite{Huber2011}.  In the context of human mobility, one of the signatures of this bias is the accentuated increase in short-term re-visitation probabilities~\cite{Barbosa2015}. 

To capture this particular facet, for a given user $u$ visiting location $\ell$, we define the number of unique intermediate locations between visits $i$ and $i+1$ to a specific location $\ell$ as $\kappa^u _{\ell,i}$. Thus, the probability of user $u$ to visit $\kappa$ unique locations before returning to a particular location, given $k_{\ell}$ total visits to that location is, 
\begin{equation}
    P_u(\kappa) = \frac{1}{\sum_{\ell} k_{\ell} - 1} \sum_{\ell} \sum_i^{k_{\ell}-1} \delta_{\kappa, \kappa^u _{\ell,i}},
    \label{eq:pkappa}
\end{equation}
with the population average $P(\kappa)$ calculated in the usual way. The quantity $P(\kappa)$ represents the distribution of lengths of subsequences of unique intermediate events, before a particular event is repeated, and reflects sequence-based biases inherent in memory recall. 

In Fig.~\ref{fig:serial_metrics}C (inset) we plot $P(\kappa)$ for the three datasets, once again finding a broad distribution. In all cases there is a significantly higher probability to visit only a few intermediate locations before returning to a particular one. In this way, the trends broadly reflect the same seen for the distribution of $P(\Delta t)$, however, unlike for inter-return times, we see no differences between the virtual and physical domains at any scale. The observed discrepancies at the intra-day range, where there were markedly more returns in Web-browsing, disappear at the event-level. Indeed, all three distributions have the same form $P(\kappa) \sim \kappa^{-\xi} \exp\left(-\kappa/\delta \right)$, with a common scaling exponent $\xi \approx 1.5$ and $\delta_{\textrm{Web}} \approx  1670$, $\delta_{\textrm{Weeplaces}} \approx 284$, $\delta_{\textrm{Brightkite:}} \approx 180$. Rescaling with respect to the cut-offs $\delta$, we find that the three distributions collapse on to a single curve.

\subsubsection{Visitation Frequency} 

Finally, we analyze the distribution of visitation frequencies, the number of occurrences of a particular location in the mobility trajectory of users. In Fig.~\ref{fig:serial_metrics}A, we plot the probability distribution $P(v)$, where $v$ represents the number of visits to a particular location $\ell$, averaged across all locations and users. 
All three datasets follow a broad power-law distribution $P(v) \sim v^{-\nu}$, 
where $\nu_{\textrm {Web}} \approx 2.1$, $\nu_{\textrm 
{Weeplaces}} \approx 2.2$ and $\nu_{\textrm {BrightKite}} \approx 2.4$. We note the presence of an exponential cut-off, $\exp(-v/\theta)$, in the physical datasets, where $\theta_{\textrm{Weeplaces}} \approx 210$ and $\theta_{\textrm{BrightKite}} \approx 348$. This likely stems from the comparatively fewer unique locations and the cost associated with physical accessibility in the offline space, as compared to the virtually limitless and cost-free browsing in the online space. Yet, aside from the cut-off, the similar scaling behavior implies that users attribute their relative time and attention to locations in common ways regardless of whether they are physical or virtual locations. 

In the physical context, similar scaling results have been observed in a variety of mobile-phone traces~\cite{Gonzalez2008a} and trajectories inferred from public transportation~\cite{Hasan2013a,Zheng2017}. In the virtual context, it has been observed in online video games~\cite{Szell2012} and online shopping behavior~\cite{Zhao_2016}. The simultaneous tracking of online and offline movement of smartphone users, revealed a strong correlation in the respective visitation frequencies~\cite{Zhao2014}, further indicating that the two types of movement are governed by a common mechanism.
 
\subsection{Model} 

\subsubsection{Motivation}

The empirical results suggest that the statistical patterns of time-independent metrics in both physical and virtual movement are strikingly similar. Given this observation, we next seek to model the underlying generative processes behind this phenomenon.  

One of the leading models capturing mobility in the physical space is based on exploration and preferential-return (EPR)~\cite{Song2010}, a phenomenological model that takes as input, empirical distributions of displacement and waiting-times. The essential ingredients of the model are decisions based on whether to explore a new location, or return to a previously visited location as a linear function of prior visits. In addition there are saturation effects, in the sense that the decision to visit new locations is an inverse function of the number of locations already visited. The model reproduces several empirically measured mobility features such as temporal discovery rates, visitation frequencies, and the phenomenon of ultraslow diffusion. While the locations are essentially statistically identical (their specific attributes or relevance in terms of mobility is not considered), the physical distance plays a \emph{symmetry-breaking} role in terms of transitions between locations closer in space being favored over those much further away. The model has seen several extensions, including visits to locations being based on a \emph{popularity measure} as opposed to frequency of prior visits~\cite{Hasan2013a}, as well as incorporating the gravity model~\cite{erlander_1990_gravity} to partition the population into \emph{returners} and \emph{explorers} based on their mobility profiles~\cite{Pappalardo2015}. Incorporating an additional decision step to revisit recent locations in the mobility trajectory, also reproduces aspects of the recency effect~\cite{Barbosa2015}. The EPR model has also been deployed to study virtual mobility based on visits between reddit forums. Here forums were clustered into topical communities and transits between communities was modeled in terms of topical relevance, in analogy to a \emph{popularity measure}~\cite{Hu_2019}.

However, while these family of models are designed to reproduce the empirical data well, being phenomenological in nature, they do not capture the generative processes behind physical and virtual mobility. By leveraging on empirically observed inputs, and directly building in choice mechanisms, such as preferential return or sampling popular locations, the models are unable to account for the rich variety of behaviors found in more general studies of human foraging and decision making. There is then a need for models that are capable of recreating empirical mobility trends from simple, intuitive principles, shedding light on the mechanisms behind human mobility while being flexible in terms of incorporating the observed diversity in human search and decision making.

For example, a study of hunter-gatherers revealed that human foraging patterns had individuals adapting their strategy to the inferred layout, or utility of resources, localizing their searches when they believed resources to be clustered \cite{Hahnloser2011}. In addition, there is a rich literature on understanding human decision making; one approach focuses on the input and output of the process (structural approach), while another emphasizes analysis of the intermediary steps (process tracing). An algorithm has been developed that uses a combination of these paradigms to effectively classify decision making strategies in an empirically observed population, demonstrating the importance of including the full spectrum of featured inherent in the decision making process, as well as the heterogeneities in the underlying strategies \cite{Riedl2008}. 

These studies suggest, that an alternative and more fundamental way to model human mobility---which is essentially a a decision-making process---is to consider a population that optimizes its gain or benefit, in response to costs associated with that particular decision. In the context of mobility, this would represent weighing locations against each other based on their relative utility to the individual before making a decision to move. This lends itself well to a Markov Chain Monte Carlo (MCMC) approach of which the Metropolis-Hastings algorithm is a particularly relevant variant~\cite{Chib2012}. Indeed, Metropolis-Hastings style models have already been employed in understanding human mobility in the context of residents of a city optimizing their travel patterns in response to air pollution~\cite{Schlink2010}. In the context of decision making, a similar framework was used to model the the Remote Associations Test (RAT), designed to test a subject's ability to solve search problems. Subjects were presented with sets of words and asked to generate additional words connecting the prompts into a cogent cognitive theme. It has been demonstrated, that a model using a Metropolis-Hastings transition scheme navigated the semantic network, constructed from the responses in a way that agreed with empirical results \cite{Bourgin2014}. The Metropolis-Hastings framework has therefore seen success in modeling human mobility in response to external stimuli, as well as cognitive search, suggesting that it is a reasonable choice for generating both general mobility trajectories and capturing human decision making processes.

\begin{figure}[t!]
\includegraphics[width=\columnwidth]{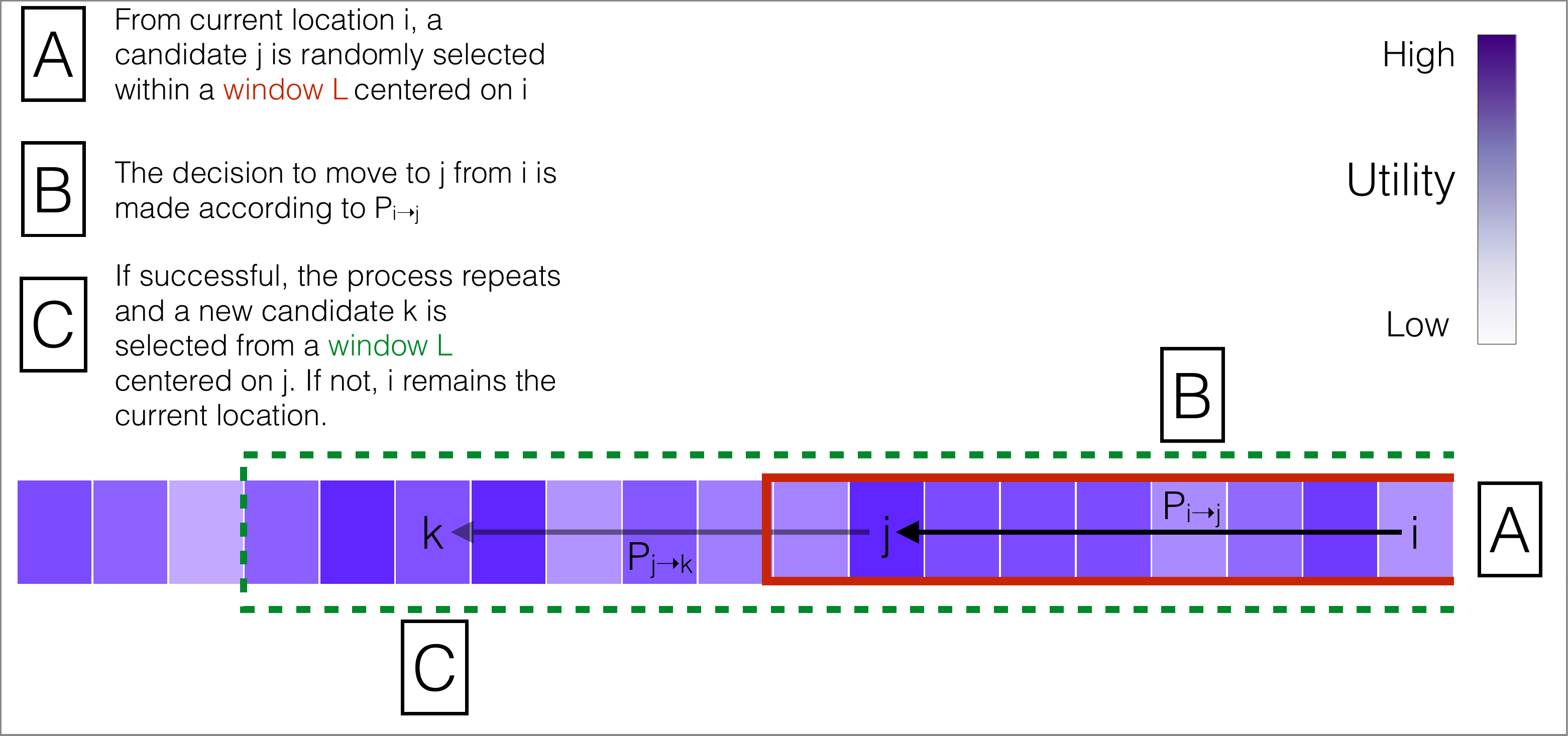}
  \caption{\textbf{Model Design} A schematic of the Metropolis-Hastings sampling procedure. The full set of locations is represented as an array of size $N_{\ell}$. Initially a location $i$ is chosen uniformly at random from $N_{\ell}$. (\textbf{A}) A candidate location $j$ is selected from within a window of length $L$ centered on the current location $i$. (\textbf{B}) The transition between $i \rightarrow j$ is governed by the transition state probability $P_{i \rightarrow j}$ (Eq.~\eqref{eq:mh}). (\textbf{C}) If $j$ is accepted, it then becomes the current location, and the window is re-centered around $j$. The process is then repeated ad infinitum. In the example above, the locations are colored according to the value of their utility. }
     \label{fig:model_design}
\end{figure}

\subsubsection{Model Design}

In our proposed model, a set of $N_u$ individuals seek to sample $N_{\ell}$ locations as a function of the utility or benefit $U_{\ell}$ of that location to the individual, balanced against an associated cost to move to that location. The utility here can be interpreted as content or relevance, in the case of websites, or features such as proximity, accessibility and attractiveness in the case of physical locations. It serves as an abstraction, covering a series of generalized attributes associated with a physical or virtual location. While we have a large degree of freedom in how we choose the utilities for locations, $U_{\ell}$, we take inspiration from prior microeconomic models, where abstract utility functions that include features such as consumption, aspiration, revenue and rate of profit, among others, take on an exponential form, $P(U_{\ell}) \sim \exp\left(-\lambda U_{\ell}\right)$, akin to the Cobbs-Douglas utility function~\cite{Gotts_2003,Xiaoping_2006, Zellner_2008}.

Next, we focus on how individuals move between locations. In models of decision making processes, while individuals are assumed to make rational choices, they are often constrained by limited information or the inability to fully exploit that information---so-called bounded rationality. Furthermore, they often seek satisfactory rather than optimal utility, in the sense of ease of making decisions, or in the case of limited resources, when the most optimal solution is not available~\cite{An_2012}. To reflect this, we model the mobility process as transitions between states $i \rightarrow j$ (representing locations) where the decision to move to a location $j$ from the current location $i$, is based on the relative differences in the utilities of the corresponding locations. The sampling of locations is conducted via the Metropolis-Hasting algorithm, with the following transition probabilities,
\begin{equation}
P_{i \rightarrow j} = 
\begin{cases}
1 &\mbox{if}\quad U_j \geq U_i \\
\exp\left[- \left(U_i - U_j \right)/T\right] &\mbox{if}\quad U_j  < U_i,
\end{cases}
\label{eq:mh}
\end{equation}
where $T$ is a \emph{temperature} variable. Eq.~\eqref{eq:mh} describes a process, where an individual necessarily moves to a location of higher utility, but weighs options in case the location has a lower utility. The extent to which an individual is willing to travel to such a location is controlled by $T$, which can be interpreted as a \emph{exploration} parameter; for $T = 0$, all transitions are made only to locations of higher utility (resembling a greedy optimization), conversely for $T \rightarrow \infty$ all locations are equally likely to be sampled irrespective of their utility. Correspondingly, the ratio $\lvert \Delta U \rvert /T$, can be thought of as a generalized cost parameter, measuring the willingness of an individual to move between locations of varying utility differences. 

In terms of which location to select in order to evaluate the transitions, we represent the full set of locations as an array of size $N_{\ell}$ and select the next location $j$ uniformly at random, from a window of length $L$ centered at $i$. If a transition is made to $j$, then the new window is now centered at $j$. The finite selection window (where typically $L \ll N_{\ell}$), reflects the bounded rationality and the imperfect knowledge of the individual. It can also be thought of as a cognitive limit in terms of the upper bound for the given number of locations an individual can assess at any given instance. Indeed, recent studies of mobility processes in both the virtual and physical domains fix this value at around twenty five locations~\cite{Goncalves_2011, Alessandretti_2018}. A schematic of the procedure is shown in Fig.~\ref{fig:model_design}. 

Thus, there are three key features of the model, (i) the heterogeneity of location utilities parameterized by $\lambda$, (ii) the exploratory tendencies of individuals parameterized by $T$, and (iii) the fraction of locations considered for sampling at any given step of the process, parameterized by $L/N_{\ell}$. We note that the model is adaptable to a wide variety of scenarios, including inherent population heterogeneities. Each individual $u$, for instance, can be associated with a unique distribution of utilities $P_u(U_{\ell})$, reflecting difference in how individuals might perceive the same location. Similarly, to represent variations in exploratory tendencies, the parameter $T$ can be assigned from some distribution $P(T)$, and differences in cognitive limits can be modeled via assigning variable $L$ to each individual. In what is to follow, however, for the sake of simplicity, we assume a common set of parameters for the entire population. As outputs of our model we measure the mobility distributions in the event space $S(n)$, $P(v)$ and $P(\kappa)$.

\begin{figure}[t!]
\includegraphics[width=.85\columnwidth]{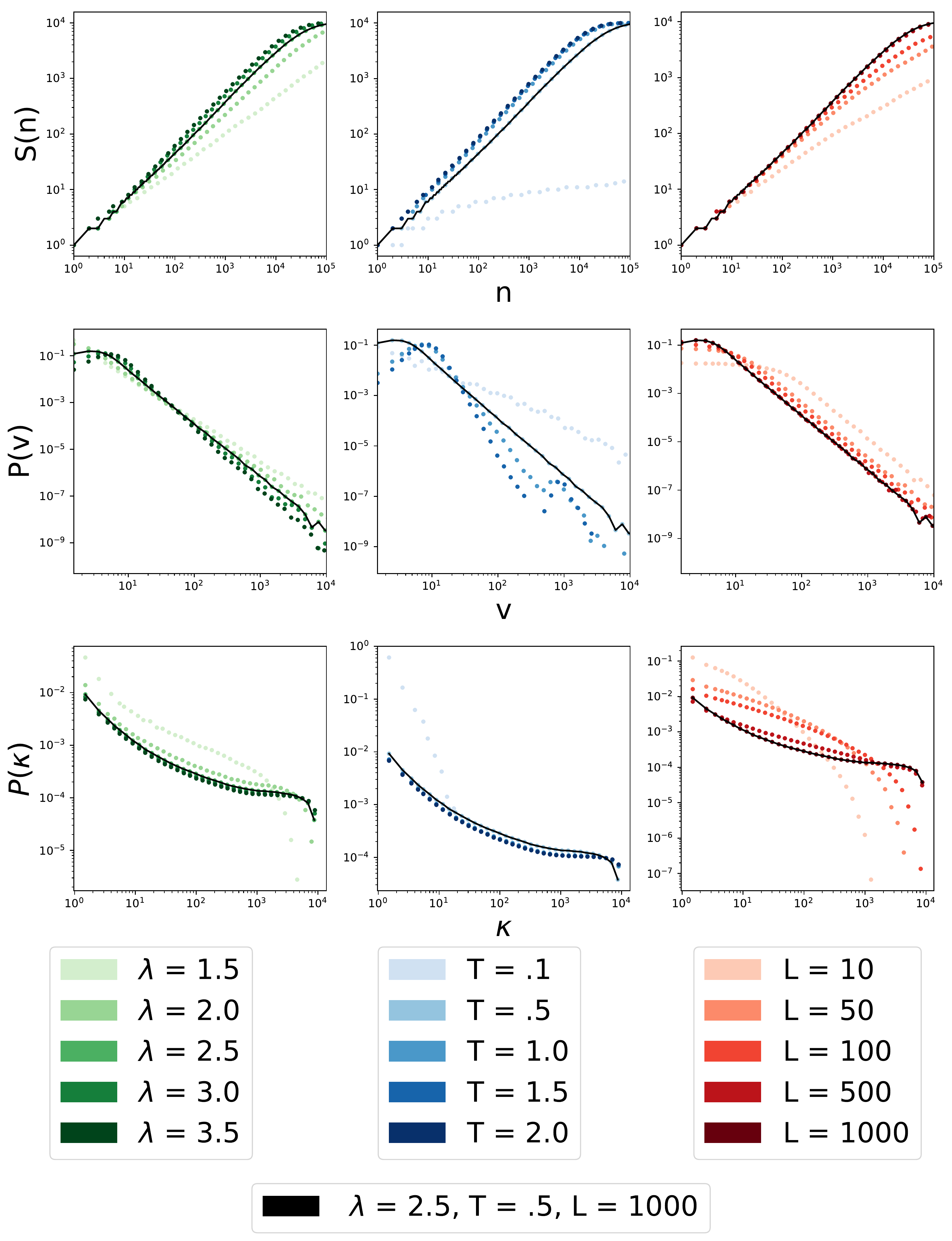}
  \caption{\textbf{Model Results} The result for simulating the Metropolis-Hasting sampling method for $N_u = 10^3$ individuals, traveling over $N_l = 10^4$ locations for $10^5$ events each. The model outputs for a range of parameters $S(n)$ top-panel, $P(\kappa)$ middle-panel and $P(v)$ bottom panel. A reference set $\lambda = 2.5$, $T = 0.5$ and $L = 10^3$ is chosen and shown as a solid black curve in each plot. Scans over each value of the parameters are shown as colored curves: green ($\lambda$), blue ($T$) and red ($L$).}
     \label{fig:model_scan}
\end{figure}

\subsubsection{Model results}

We simulate our model for $N_u = 10^3$ individuals sampling a location array of $N_{\ell}=10^4$ for $n_u = 10^5$ steps of the process. We set, $T = 0.5$, $L = 10^3$, and $\lambda = 2.5$ as a reference set, and scan across each parameter separately (while fixing the other two to the reference value), to determine its influence on the distributions of interest. We plot our results in Fig.~\ref{fig:model_scan}, showing $S(n)$ (upper-panel), $P(\kappa)$ (middle-panel) and $P(v)$ (bottom panel). The curve corresponding to the reference set of parameters is shown as a solid black line in each plot, and the curves corresponding to the scan over the parameters are shown as colored dashed lines, $\lambda$ (green), $T$ (blue), and $L$ (red). We next describe the effect of each of the parameters in our model.

\emph{Heterogeneity of utilities:} For more heterogenous distributions of $U_{\ell}$ (lower values of $\lambda$), there are those locations whose utility is well differentiated from other locations. Correspondingly, we see a lower slope in the curve for $S(n)$, indicating that location visits tend to be restricted to a fewer set of locations, with only 10\% of locations eventually discovered at the end of the simulation process. This is also reflected in a broad distribution for $P(v)$ and a rapid fall-off in $P(\kappa)$ indicating only a few intermediate locations are visited before returns. With increasing $\lambda$, and therefore lower variance in the utilities ($\sim \lambda^{-2}$), we have a more homogenous distribution of utilities that are closely spaced, and therefore favoring many more transitions between locations. This is reflected as a progressive increase in $S(n)$, a shallower distribution of $P(v)$, and finally a flattening of $P(\kappa)$, indicating a suppression of the recency effect.

\emph{Exploratory parameter:} Increasing $T$ has the same effect as lowering the variance in the utility distribution. For low values of $T$, transitions are favored only between a few locations, and thus, we see a similar behavior as in higher variance for $U_{\ell}$. However, we note the relatively rapid fall-off in $P(\kappa)$ for $T = 0.1$, indicating a very strong recency bias. However, the effect is somewhat artificial; for very small values of $T$, the process is similar to greedy optimization, where a limited set of unique locations are visited. Consequently, it follows that only few intermediate locations will be visited before returns. This is also reflected in a saturation in $S(n)$ where only .1 \% of locations are visited leading to a shallow distribution for $P(v)$. As $T$ is increased and transitions are more favored, we see all three quantities converging to a stable distribution. This is to be expected as for large values of $T$, $P_{i \rightarrow j} \approx 1$ and all locations are equally likely to be visited. The results indicate that the process is relatively insensitive to the temperature, except at low values ($T \ll 1$). 

\emph{Window size:} Decreasing the window size leads to relatively fewer locations being sampled, independent of whether the distribution of utilities has high or low variance. Consequently, we see a saturation in $S(n)$, similar to that seen for a utility distribution with higher variance, although markedly less pronounced than for the saturation seen for low $T$.  After a relatively small increase in $L$, $S(n)$ appears to converge to a distribution that eventually becomes independent of $L$. A similar effect is seen for $P(v)$, while the largest influence of $L$ is seen for the distribution of sub-sequences $P(\kappa)$, with markedly different distributions for different window sizes. It is apparent that as one increases $L$, the recency-bias gradually vanishes. 

\begin{figure}[t!]
\includegraphics[width=\columnwidth]{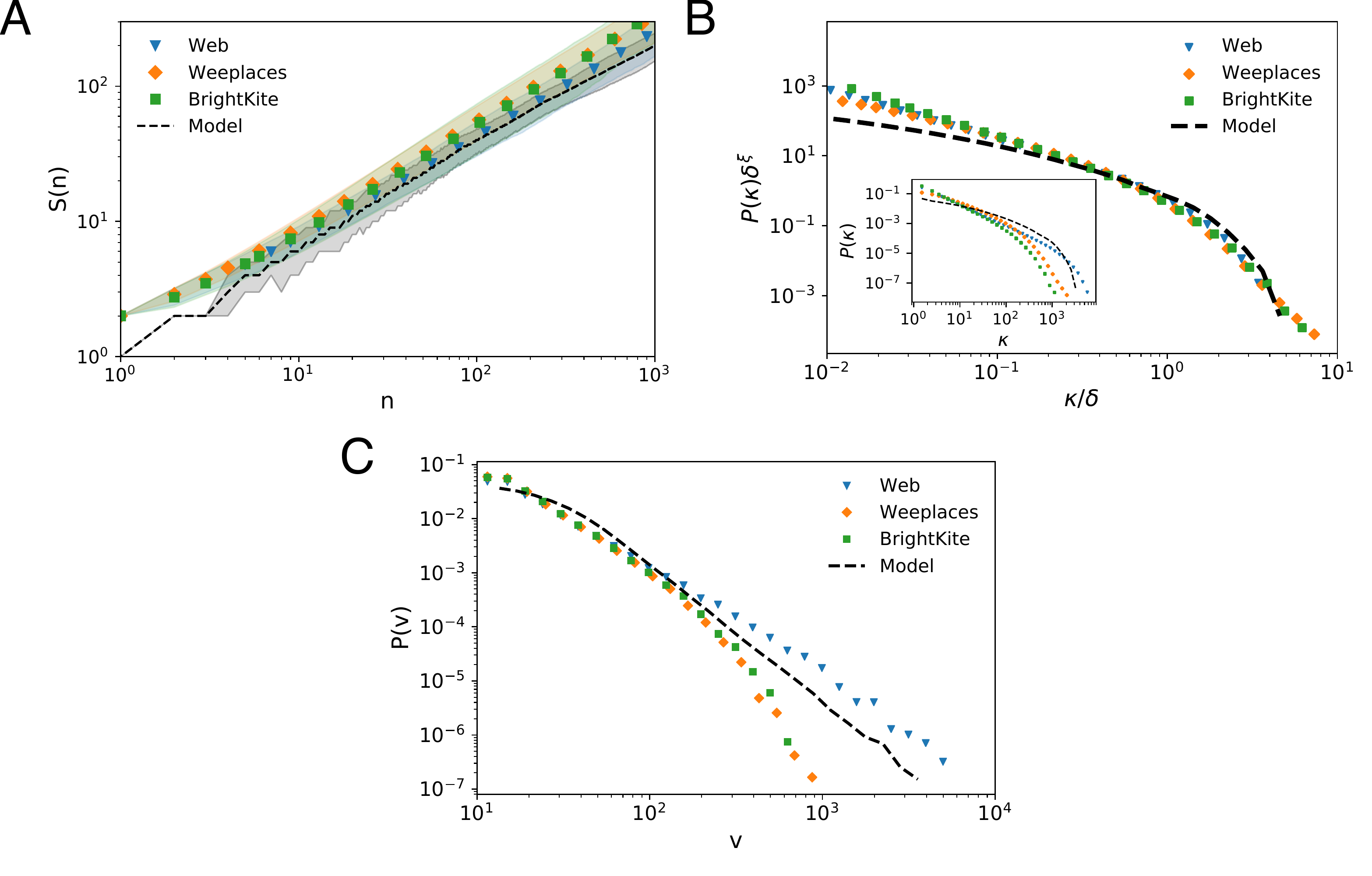}
  \caption{\textbf{Correspondence between the model and empirical data} We show the same quantities as in Fig.~\ref{fig:serial_metrics} compared to the results of the model (shown as dashed curve). The parameters used were $\lambda = 5, T = 0.3$ and $L = 26$, and the curves were generated as a result of simulating $N_u = 10^3$, $N_{\ell} = 10^4$ for $10^5$ events. In all cases we find good agreement between the model and empirical observations.}
    \label{fig:model}
\end{figure}

The results indicate that a broad spectrum of mobility patterns can be generated as a function of the interpolation between the three parameters. Indeed, some combination of these can be tuned to simultaneously generate the observed empirical trends in our dataset. As a proof-of-concept, in Fig.~\ref{fig:model}, we plot the results of our model for a set of parameters ($\lambda = 5, T = 0.3, L = 26$), that reproduces the trends seen in Fig.~\ref{fig:serial_metrics} including for $S(n)$, (Fig.~\ref{fig:serial_metrics}A), $P(\kappa)$ (Fig.~\ref{fig:serial_metrics}B) and $P(v)$ (Fig.~\ref{fig:serial_metrics}C), showing very good agreement with the data. It is instructive to note that the window-size, $L = 26$, is virtually the same as the empirically determined characteristic number of locations visited by individuals in physical movement~\cite{Alessandretti_2018}. 

\section{Discussion}

Taken together our results shed new insight on the similarities and differences between physical and virtual movement. 
The spatial constraints inherent in the former, are reflected in its differences with the latter, at short time-scales corresponding to the intra-day regime. For virtual movement, there is a relatively more rapid activity rate, faster returns to locations, and the tendency to visit a much greater number of unique locations in a comparable time frame as compared to physical movement. Yet, even at the temporal level, when moving to the inter-day regime, we find similar statistical patterns in both the inter-return and inter-event times, suggesting that individuals allocate time spent in both flavors of locations in relatively the same way. The other difference is in the number of unique locations visited over a comparable period of time, where an order of magnitude separates the locations discovered in the virtual domain as compared to the physical domain. However, the rate at which discoveries are made (as measured by the slope) is also similar. The observed agreement in the long-time statistics indicates that location-based decision making may differ when the question is ``how long will it take me to get there" (short term) but not when the question becomes ``when do I want to go" (long term.) That is, when the decision is based purely on preference and not the physical ability to move, it occurs in the same way regardless of physical or virtual space.

This becomes clear, when one moves to the time-independent space of events, or sequences of location visits. Here across three different metrics, the event discovery $S(n)$, the recency bias $P(\kappa)$ and the visitation frequency $P(v)$ the statistical distributions are virtually identical, hinting at a generalized mechanism behind both types of mobility patterns. Inspired by past research on human decision making, we proposed a model based on the Metropolis-Hastings framework, that with only three intuitive mechanisms---how locations are perceived, the decision to move based on perception, and the number of locations assessed ay any given point of time---is capable of reproducing the empirically observed  time-independent mobility statistics in both the virtual and physical domains. We emphasize that, unlike existing phenomenological models, our model does not require any empirical input, and can be applied to a range of processes involving human decision making. Two of the processes in our model are parameterized by free parameters, that can be tuned to produce a wide spectrum of behaviors, while the other, the distribution of utilities, is derived from well-founded principles in microeconomic theory. Given the agreement between the model and the data, for a window size of $L=26$, the same as the empirically observed characteristic number of locations visited by people, effectively the model has one free parameter, the temperature $T$, which as discussed earlier can be be interpreted as an exploration parameter. Our results suggest that rather than treating online and offline movement separately, it should be considered under a single framework, although many more facets of online activity must considered before coming to a strong conclusion.  Indeed, as mentioned in the introduction, an increasing amount of socioeconomic activity in the physical world is being replaced by virtual analogs. While the manuscript deals specifically with mobility, future directions will explore other facets of human activity and its similarities and differences in the virtual and physical domains.

\begin{acknowledgments}
This work was supported by the US Army Research Office under Agreement Number W911NF-18-1-0421.
\end{acknowledgments}

\def\urlprefix{}
   \def\url#1{}
 \def\eprintprefix{}
 \def\eprint#1{}

\end{document}